\documentclass{article}

\usepackage{PRIMEarxiv}
\usepackage{listings}
\usepackage{cite}
\usepackage{float}
\usepackage[utf8]{inputenc} 
\usepackage[T1]{fontenc}    
\usepackage{hyperref}       
\usepackage{url}            
\usepackage{booktabs}       
\usepackage{amsfonts}       
\usepackage{nicefrac}       
\usepackage{microtype}      
\usepackage{lipsum}
\usepackage{fancyhdr}       
\usepackage{graphicx}       
\graphicspath{{media/}}     
\usepackage{indentfirst}

\title{jointVIP: Prioritizing variables in observational study design with joint variable importance plot in R}

\author{
  Lauren D. Liao \\
  Division of Biostatistics \\
  University of California, Berkeley \\
  \texttt{ldliao@berkeley.edu} \\
   \And
  Samuel D. Pimentel \\
  Department of Statistics \\
  University of California, Berkeley \\
  \texttt{spi@berkeley.edu} \\
}

\begin{document}
\setlength\parindent{24pt}\setlength{\parskip}{0.0pt plus 1.0pt}
\maketitle

\begin{abstract}
Credible causal effect estimation requires treated subjects and controls to be otherwise similar. In observational settings, such as analysis of electronic health records, this is not guaranteed. Investigators must balance background variables so they are similar in treated and control groups. Common approaches include matching (grouping individuals into small homogeneous sets) or weighting (upweighting or downweighting individuals) to create similar profiles. However, creating identical distributions may be impossible if many variables are measured, and not all variables are of equal importance to the outcome. The joint variable importance plot (\texttt{jointVIP}) package to guides decisions about which variables to prioritize for adjustment by quantifying and visualizing each variable's relationship to both treatment and outcome.
\end{abstract}


\section{Statement of need}
Consider an observational study to measure the effect of a binary treatment variable (treated/control) on an outcome, in which additional covariates (background variables) are measured. A covariate may be associated with outcomes, and it may also differ in distribution between treated and controls; if the covariate is associated with treatment and outcome, the covariate in question is a confounder. Ignored confounders introduce bias into treatment effect estimates. For instance, when testing a blood pressure drug, if older patients both take the drug more and have worse initial blood pressure, a simple difference in mean blood pressure between treated and control subjects will understate the drug's benefits. Confounding can be addressed by matching, under which blood pressure is compared only within pairs of patients with similar ages, or by weighting, in which older control subjects receive larger weights than younger control subjects when averaging blood pressure. When many potential confounders are measured, however, neither matching nor weighting can perfectly adjust for all differences, and researchers must select which variables to focus on balancing.

Current practice for selecting variables for adjustment focuses primarily on understanding the treatment relationship, via tools such as balance tables and the Love plot \cite{ahmed2006, greifer2021, hansenbowers2008, rosenbuam1985, stuart2011}. A key metric is the standardized mean difference (SMD), or the difference in treated and control means over a covariate measure in standard deviations. Researchers commonly try to adjust so that all SMD values are moderately small, or focus on adjustments for variables with the largest initial SMD. However, these approaches neglect important information about the relationship of each covariate with the outcome variable, which substantially influences the degree of bias incurred by ignoring it.

To improve observational study design, we propose the joint variable importance plot (jointVIP) \cite{liao2023}, implemented in the \texttt{jointVIP} package. The jointVIP represents both treatment and outcome relationships for each variable in a single image: each variable's SMD is plotted against an outcome correlation measure (computed in a pilot control sample to avoid bias from multiple use of outcome data). Bias curves based on unadjusted, simple one-variable omitted variable bias models are plotted to improve variable comparison. The jointVIP provides valuable insight into variable importance and can be used to specify key parameters in existing matching and weighting methods.

\newpage
\section{Development}
The \texttt{jointVIP} package was created in the R programming language \cite{cran}. The package uses the S3 object system and leverages system generic functions, \texttt{print()}, \texttt{summary()}, and \texttt{plot()}. Plotting the jointVIP object outputs a plot of the \texttt{ggplot2} class. An interactive R Shiny application, available online at https://ldliao.shinyapps.io/jointVIP/, showcases the package.

\section{Usage}

The \texttt{jointVIP} package is available from the Comprehensive R Archive Network \href{https://CRAN.R-project.org/package=jointVIP}{CRAN} and \href{https://github.com/ldliao/jointVIP}{GitHub}. To create an object of the jointVIP class, the user needs to supply two datasets and specify the treatment, outcome, and background variable names. Two processed datasets, ``pilot'' and ``analysis'' samples, are in the form of data.frames. The analysis sample contains both treated and control groups. The pilot sample contains only control individuals, and they are excluded from the subsequent analysis stage. The treatment variable must be binary: 0 specified for the control group and 1 specified for the treated group. Background variables are measured before both treatment and outcome. The outcome of interest can be either binary or continuous.

\begin{lstlisting}
# installation using CRAN:
# install.packages("jointVIP")

# installation of development version using GitHub:
# remotes::install_github('ldliao/jointVIP') 

library(jointVIP)
\end{lstlisting}

We demonstrate the utility of this package to investigate the effect of a job training program on earnings \cite{causaldata, dehejia1999causal, lalonde1986evaluating}.  The treatment is whether the individual is selected for the job training program. The outcomeis earnings in 1978. Covariates are age, education, race/ethnicity, and previous earnings in 1974 and 1975.  After \href{https://cran.rstudio.com/web/packages/jointVIP/vignettes/jointVIP.html}{preprocessing both dataset and log-transforming the earnings}, we use the \texttt{create\_jointVIP()} function to create a jointVIP object stored as \texttt{new\_jointVIP}.

\begin{lstlisting}
# first define and get pilot_df and analysis_df
# they should both be data.frame objects

treatment <- "treat"
outcome <- "log_re78"
covariates <- c("age", "educ", "black", 
                "hisp", "marr", "nodegree",
                "log_re74", "log_re75")

new_jointVIP = create_jointVIP(treatment = treatment,
                               outcome = outcome,
                               covariates = covariates,
                               pilot_df = pilot_df,
                               analysis_df = analysis_df)
                               
\end{lstlisting}

\begin{figure}[H]
  \centering
  \includegraphics[width=0.75\textwidth]{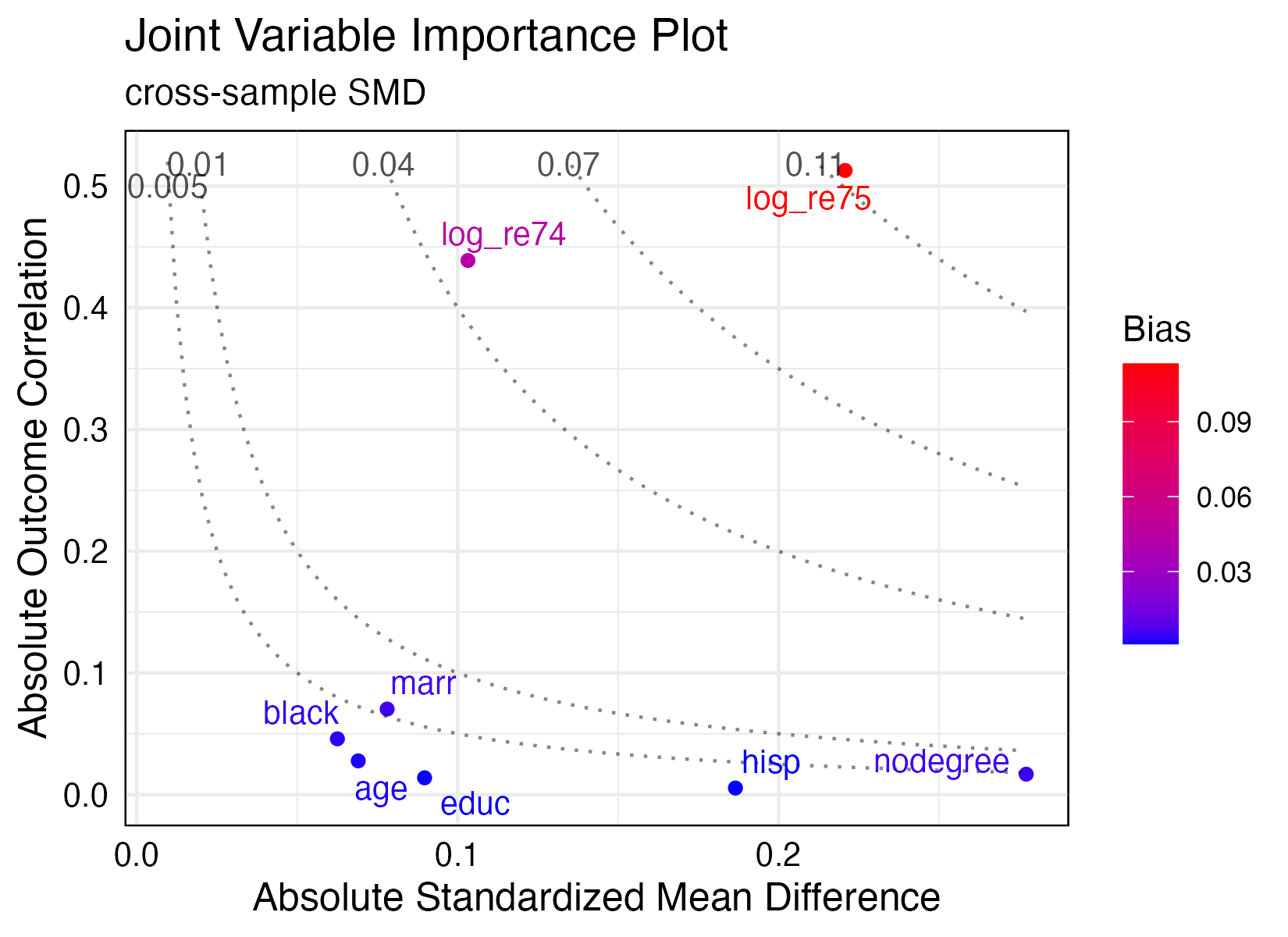}
  \caption{Joint variable importance plot example}
  \label{fig:fig1}
\end{figure}

The \texttt{plot()} function displays a jointVIP (Figure \ref{fig:fig1}). The x-axis describes treatment imbalance in SMD (computed with a denominator based on the pilot sample as in \cite{liao2023}). The y-axis describes outcome correlations in the pilot sample. The \texttt{summary()} function outputs the maximum absolute bias and the number of variables required for adjustment above the absolute bias tolerance, \texttt{bias\_tol}. The \texttt{bias\_tol} parameter can be used in the \texttt{print()} function to see which variables are above the desired tolerance. Additional tuning parameters can be specified in these functions, for details and examples, see \href{https://cran.rstudio.com/web/packages/jointVIP/vignettes/additional_options.html}{the additional options vignette}.

\newpage
\begin{lstlisting}
plot(new_jointVIP)
\end{lstlisting}

\begin{lstlisting}
summary(new_jointVIP, 
        smd = "cross-sample",
        use_abs = TRUE,
        bias_tol = 0.01)
        
# > Max absolute bias is 0.113
# > 2 variables are above the desired 0.01 absolute bias tolerance
# > 8 variables can be plotted

print(new_jointVIP,
      smd = "cross-sample",
      use_abs = TRUE,
      bias_tol = 0.01)
      
# >           bias
# > log_re75 0.113
# > log_re74 0.045
\end{lstlisting}

To interpret our working example, the most important variables are the previous earning variables in 1975 and 1974, \texttt{log\_re75} and \texttt{log\_re74} variables, respectively. Using the traditional visualization method, the Love plot, would only identify variables based on the SMD. The same information can be interpreted from the x-axis of the jointVIP. For example, the Love plot would indicate variables, \texttt{nodegree} and \texttt{hisp}, to be more important for adjustment than \texttt{log\_re74}. In comparison, those variables, \texttt{nodegree} and \texttt{hisp}, show low bias using the jointVIP.

\begin{figure}[H]
  \centering
  \includegraphics[width=0.75\textwidth]{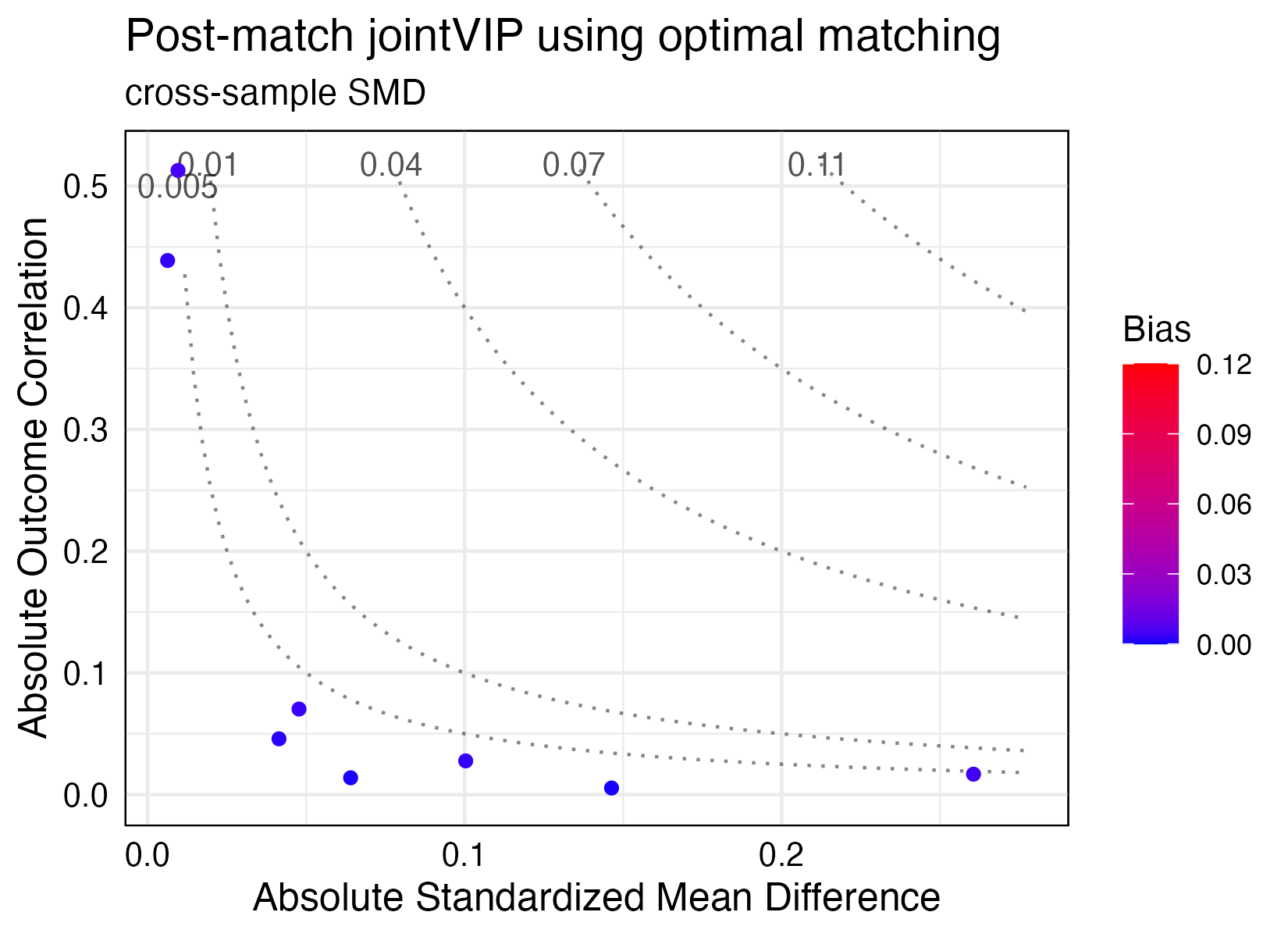}
  \caption{Post match example showing balanced sample based on new mean differences.}
  \label{fig:fig2}
\end{figure}

After adjusting for variables, for example, using optimal matching \cite{optmatch, stuart2011} to select pairs for analysis, a post-adjustment dataset, \texttt{post\_analysis\_df}, can be used to create a post adjustment object of class \texttt{post\_jointVIP}. The \texttt{create\_post\_jointVIP()} function can be used to visualize and summarize the post adjustment results, as seen in Figure \ref{fig:fig2}. The functions: \texttt{summary()}, \texttt{print()}, and \texttt{plot()} all can take in the post\_jointVIP object and provide comparison between original and post adjusted jointVIPs.

\newpage 
\begin{lstlisting}
post_optmatch_jointVIP <- create_post_jointVIP(new_jointVIP, 
                                               post_analysis_df = optmatch_df)

plot(post_optmatch_jointVIP, 
     plot_title = "Post-match jointVIP using optimal matching")
\end{lstlisting}

\begin{lstlisting}
summary(post_optmatch_jointVIP)

# > Max absolute bias is 0.113
# > 2 variables are above the desired 0.01 absolute bias tolerance
# > 8 variables can be plotted
# >
# > Max absolute post-bias is 0.005
# > Post-measure has 0 variable(s) above the desired 0.005 absolute bias tolerance

print(post_optmatch_jointVIP)

# >           bias post_bias
# > log_re75 0.113     0.005
# > log_re74 0.045     0.003
\end{lstlisting}

\section{Discussion}
We have developed user-friendly software to prioritize variables for adjustment in observational studies. This package can help identify important variables related to both treatment and outcome. One limitation is that each background variable is individually evaluated for bias. Thus, conditional relationships, interactions, or higher moments of variables need to be carefully considered or preprocessed by the user.

\section*{Acknowledgments}

The authors thank Emily Z. Wang for helpful comments.
SDP is supported by Hellman Family Fellowship and by the National Science Foundation (grant 2142146). LDL is supported by National Science Foundation Graduate Research Fellowship (grant DGE 2146752).

\bibliographystyle{unsrt}  
\bibliography{references}

\end{document}